\documentclass[twocolumn,a4paper,reprint,amssymb,aps,prd,groupedaddress,superscriptaddress]{revtex4-1}
\pdfoutput=1
\usepackage{dcolumn}
\usepackage[left=1.5cm,top=2cm,right=1.5cm,bottom=2cm]{geometry}
\usepackage{array}
\usepackage{xcolor}
\usepackage{amsmath}
\usepackage{graphicx,bm,color}
\usepackage{subfigure,bm}
\usepackage{amsfonts}
\usepackage{lipsum}
\usepackage{enumitem}  
\usepackage{booktabs}
\usepackage{array}
\usepackage{hyperref} 
\usepackage{changes}

\begin{document}
\title{Remedy of some cosmological tensions via effective phantom-like behavior of interacting vacuum energy}

\author{Suresh Kumar}
\email{suresh.math@igu.ac.in}
\affiliation{Department of Mathematics, Indira Gandhi University, Meerpur, Haryana-122502, India}
\affiliation{Department of Mathematics, National Institute of Technology, Kurukshetra, Haryana-136119, India}
\affiliation{Department of Mathematics, BITS Pilani, Pilani Campus, Rajasthan-333031, India}

\begin{abstract}
Since physics of the dark sector components of the Universe is not yet well-understood, the phenomenological studies of non-minimal interaction in the dark sector could possibly pave the way to theoretical and experimental progress in this direction. Therefore, in this work, we intend to explore some features and consequences of a phenomenological interaction in the dark sector. We use the Planck 2018, BAO, JLA, KiDS and HST data to investigate two extensions of the base $\Lambda$CDM model, viz., (i) we allow the interaction among vacuum energy and dark matter, namely the I$\Lambda$CDM model, wherein the interaction strength is proportional to the vacuum energy density and expansion rate of the Universe, and (ii) the I$\Lambda$CDM scenario with free effective neutrino mass and number, namely the $\nu$I$\Lambda$CDM model. We also present comparative analyses of the interaction models with the companion models, namely, $\Lambda$CDM, $\nu\Lambda$CDM, $w$CDM and $\nu w$CDM. In both the interaction models, we find non-zero coupling in the dark sector up to 99\% CL with energy transfer from dark matter to vacuum energy, and observe a phantom-like behavior of the effective dark energy without actual ``phantom crossing". The well-known tensions on the cosmological parameters $H_0$ and $\sigma_8$, prevailing within the $\Lambda$CDM cosmology, are relaxed significantly in these models wherein the $\nu$I$\Lambda$CDM model shows consistency with the standard effective neutrino mass and number. Both the interaction models find a better fit to the combined data compared to the companion models under consideration.

\end{abstract}

\maketitle

\section{INTRODUCTION}
The standard model of the modern cosmology, namely the $\Lambda$CDM (cosmological constant $\Lambda$ + cold dark matter) model, exhibits an excellent agreement with variety of observational data such as the cosmic microwave background (CMB) observations from Planck \cite{Planck2018}, baryonic acoustic oscillations (BAO) \cite{BAO}, Supernovae type Ia (SNe) \cite{SNIe} are to mention a few. Despite the great success of the $\Lambda$CDM model in representing the accelerated expansion and large scale structure (LSS) of the Universe in line with the observations, there are some potential and compelling problems with this model which motivate the investigation of its extension or alternative cosmological models. For, a  profound problem is the well-known ``cosmological constant fine tuning problem", which refers to the fact that most quantum field theories predict a huge
cosmological constant with energy density $\rho_{\Lambda}\sim 10^{71}$ GeV$^{4}$ from the energy of the quantum vacuum at present, which is about 118 orders of
magnitude larger than the cosmological upper bounds yielding $\rho_{\Lambda}\sim 10^{-47}$GeV$^{4}$, leading to an
extreme fine-tuning problem. Among the problems in representing the small scale structures in the Universe encountered by the $\Lambda$CDM model, there exists the ``missing satellite problem" which refers to the discrepancy between the predicted number of
subhalos in N-body simulations and the observed number of subhalos. It is further complicated by the ``Too Big To Fail" problem because the $\Lambda$CDM predicted satellites are too massive and dense compared to the observed ones (see \cite{rev17a,rev17b} for recent reviews and references therein). 

Further, recent measurements of the present Hubble constant $H_0$  are in serious disagreement with the ones inferred from the Planck CMB experiments within the $\Lambda$CDM cosmology. To be more precise, the Hubble Space Telescope (HST) observations of the Large Magellanic Cloud Cepheids yield the measurement (R19): $H_0 = 74.03 \pm 1.42$  km s${}^{-1}$ Mpc${}^{-1}$ which finds 4.4$\sigma$ level tension with Planck CMB $\Lambda$CDM value: $H_0 = 67.4 \pm 0.5$  km s${}^{-1}$ Mpc${}^{-1}$, raising the discrepancy beyond a plausible level of chance \cite{R19}. The measurements from other approaches also prefer higher values of $H_0$ showing tensions between 4$\sigma$ to $5.8\sigma$ when compared to the $H_0$ value derived from Planck CMB $\Lambda$CDM \cite{verde19}. Thus, $H_0$ tension seems a compelling and potential tension in the $\Lambda$CDM cosmology. Here, it deserves mention that the calibration using methods not involving Cepheid stars (as in R19) find tension with the R19 results. In particular, with the calibration using the Tip of the Red Giant Branch (TRGB) \cite{freedman19}, the tension is $\sim 2\sigma$ and this inconsistency can be traced back to a tension in the distances to common SNe hosts \cite{H1}. In \cite{H1}, it has also been shown that a 0.14 shift in magnitude in the SNe absolute calibration, $M_B$, can easily solve the $H_0$ tension without requiring any additional physics. This result is in agreement with the results of \cite{H2}, where SNe are calibrated using BAO and cosmic chronometers, and that of \cite{H3} where the systematics in the dust extinction
model used in Cepheid calibration are explored.  

The measurements  of the r.m.s. fluctuation of density perturbation at $8 h^{-1}$ Mpc scale, characterized by $\sigma_8$ or the parameter $S_8= \sigma_8\sqrt{\Omega_m/0.3}$, from the cosmic shear surveys also show inconsistency at a significant level with the one given by Planck CMB data within the $\Lambda$CDM cosmology. For instance, the recent cosmic shear analysis of the fourth data release of the Kilo-Degree Survey (KiDS-1000) has reported the value $S_8=0.759^{+0.024}_{-0.021}$, which is in 3$\sigma$ tension with the prediction of the Planck CMB $\Lambda$CDM cosmology. For more details, we refer the reader to \cite{asgari2020}, and references therein, where one can infer that $S_8$ does not fully capture the degeneracy direction in all the analyses. In fact, there are varying degrees of $S_8$ tension from different surveys whereas a combined analysis from KiDS and GAMA galaxy clustering leaves no tension on $S_8$ \cite{edo2017}. So the $S_8$ tension may not be a well defined compared to the $H_0$ tension possibly due to the large systematics in the measurements from the cosmic shear surveys. Also, in the Dark Energy Survey (DES) Year 3 Results \cite{DES3}, they did not find any tension with Planck significantly undermining the validity of the $S_8$ tension.   Nonetheless, various cosmic surveys predict lower values of $S_8$ compared to the Planck CMB prediction. 

In the $\Lambda$CDM model, dark matter (DM) and dark energy (DE) are assumed to behave as separate fluids without any interactions beyond the gravitational ones. However, DM and DE are the major energy ingredients in the energy budget of the Universe, and a non-minimal interaction among these two sources of energy could be a natural possibility. There are numerous studies in the literature where the interaction between DM and DE is studied in different contexts \cite{1,2,3,4,6,8,9,11,12,13,14,15,16,17,18,19,20,21,22,23,24,25,26,27,28,29,30,31,32,33,
34,35,36,37,38,39,40,41,42,43,44,45,46,47,48,49,50,51,52,54,55,56,58,59,60,61,62,63,
64,65,66,67,68,69,70,71,72,73,74,75,76,77,78,78a,79,80,81,82,83,84,86,86a,87,88,89,90,91,93,94,
95,96,97,98,99,100,101,102,103,104,105,106,107,108,109,109a,110,111,112,113,114,115,116,sk19,118,
119,120,121,122,123,124,125,126,n1,d2,n2,n3}. It is argued that the above mentioned $H_0$ and $S_8$ tensions could possibly due to systematics in the data or some physics beyond the standard $\Lambda$CDM model \cite{issue01,issue02}. Considering the possibility of new physics, it is shown in the literature that the tensions on $H_0$ and $S_8$ can be alleviated by allowing the interaction among vacuum energy and DM wherein the interaction strength is proportional to the vacuum energy density and expansion rate of the Universe (see \cite{sk19, 95, 122, 126,n2} for some recent studies, and references therein). In the present study, we reassess this model, and in addition, we investigate constraints on the effective neutrino mass and neutrino number within this interacting vacuum energy scenario with the observational data (see Section \ref{sec3}) wherein the motivation for studying effective neutrino mass and neutrino number is deferred to the next section. All the models considered in this study are investigated recently in \cite{122}. The sole and main objective of the present study is to explore and highlight some interesting physics and features of the interaction in the dark sector of the Universe via the phenomenological model of interaction between vacuum energy and DM while providing improved observational constraints on the models compared to \cite{122}. The mathematical details of the models are described in Section \ref{sec2}. Observational constraints and discussions are presented in Section \ref{sec3} while the results are compared with some previous studies in Section \ref{sec4}. Finally, findings of this study are summarized in Section \ref{sec5}.

\section{Models of interaction in the dark sector} \label{sec2}
The expansion rate of the Universe is given by Hubble parameter $H= \dot{a}/a$, where $a$ is scale factor and an overdot denote cosmic time derivative. We assume a spatially-flat Friedmann-Robertson-Walker Universe filled with photons, neutrinos, baryons, cold DM and DE. In such a Universe, the evolution of Hubble parameter is given by the Friedmann equation:

\begin{align}
\label{friedmann}
3 H^2 = 8 \pi G (\rho_{\gamma} + \rho_{\nu} + \rho_{\rm b} + \rho_{\rm dm} +
\rho_{\rm de}  ),
\end{align}
where $\rho_{\gamma}$, $\rho_{\nu}$, $\rho_{\rm b}$,
$\rho_{\rm dm}$, and $\rho_{\rm de}$ are the energy densities of photons, neutrinos, baryons, cold DM and DE, 
respectively. We allow the interaction between DM and DE via a coupling function $Q$ given by
\begin{equation}\label{eq1}
\dot{\rho}_{\rm dm} + 3H\rho_{\rm dm} =Q= -\dot{\rho}_{\rm de} -
3H \rho_{\rm de} (1 + w_{\rm de}).    
\end{equation}
Following our previous work \cite{sk19}, we assume that the coupling term is proportional to the Hubble parameter and DE density, that is, $Q \propto H \rho_{\rm de}$. Thus, we use the following form of $Q$:
\begin{equation}\label{eq2}
Q =\xi H \rho_{\rm de},    
\end{equation}
where $\xi$ is the coupling parameter that characterizes the coupling strength between DM and DE. Solving \eqref{eq1} and \eqref{eq2}, the energy densities of DM and DE are obtained as 
\begin{equation}
\rho_{\rm dm}=\rho_{\rm dm0}a^{-3}+\frac{\xi\rho_{\rm de0}  }
{3w_{\rm de}+\xi}\Big[a^{-3}-a^{-3(1+w_{\rm de})-\xi}\Big],
\end{equation}
\begin{equation}
\rho_{\rm de}=\rho_{\rm de0} a^{-3(1+w_{\rm de})-\xi}.
\end{equation}
We see that $\xi=0$ (no interaction) yields the standard evolution of DM and DE, as expected. Further, we notice that $\xi<0$ implies energy transfer from DM to DE while $\xi>0$ implies the opposite. In this work, we consider the vacuum energy characterized by the equation of state (EoS) parameter: $w_{\rm de}=-1$. Then the effective EoS parameters of DM and DE read as follows \cite{18}:

\begin{equation}\label{eq6}
w_{\rm dm}^{\rm eff}=-\frac{\xi\rho_{\rm de}}{3\rho_{\rm dm}}\;,
\end{equation}
\begin{equation}\label{eq8}
w_{\rm de}^{\rm eff}=-1+\frac{\xi}{3}\;.
\end{equation}

The present Universe may well be dominated by a phantom-like behavior of DE. Such a behavior can be explained by invoking a phantom field. However, a phantom field suffers from some undesirable problems such as the instabilities and  violation of the null energy condition. So, it could be interesting to realize an effective phantom-like behavior without introducing a phantom field. Note that effective phantom-like behavior of DE means that
the effective DE energy density is positive and increases with time, wherein the effective
EoS parameter of DE stays less than $-1$. A phantom-like behavior of DE, without invoking any phantom fields, is studied in the normal branch of the DGP cosmological solution in \cite{ep1,ep2,ep3,ep4,ep5}. Phantom behavior via cosmological creation of particles is studied in \cite{ep6}. From eq.(\ref{eq8}), we notice that the condition for effective phantom behavior of DE is $\xi<0$. Thus, it is readily possible to realize phantom-like behavior in the framework of interacting vacuum energy model.

As described in \cite{Planck2015}, the neutrino flavour oscillation experiments allow normal mass hierarchy with the minimal mass $ \sum m_{\nu}= 0.06$ eV. On the other hand, many neutrino mass models are consistent with current observations, and one does not find strong preference over the others with compelling theoretical reasons. So considering the fact that the $\sum m_{\nu}$ is non-zero, extension of the base model with free $\sum m_{\nu}$ is naturally a well-motivated extension
of the base model. Next, many extensions of the Standard
Model of particle physics allow the existence of new light particles while it is usual to parametrize the dark radiation density in the early Universe by $N_{\rm eff}$, viz., 
well after electron-positron annihilation the total relativistic energy density in neutrinos and any other dark radiation reads
\begin{equation}
\rho_\nu=N_{\rm eff}\dfrac{7}{8}\left(\dfrac{4}{11}\right)^{4/3}\rho_\gamma.
\end{equation}
The standard model of particle physics predicts, $N_{\rm eff}=3.046$, and it corresponds to two massless and one massive neutrino. Note that an additional increment of 1 in $N_{\rm eff}$ could well correspond to a fully thermalized sterile neutrino which  decoupled at $T \lesssim 100$ MeV. So hereafter we will refer $\sum m_{\nu}$ and $N_{\rm eff}$ to as effective neutrino mass and number, respectively. In this work, we intend to investigate the effects of the interaction in the dark sector on $\sum m_{\nu}$ and $N_{\rm eff}$. In summary, we investigate the following two models in this work. \\

\noindent\textbf{I$\Lambda$CDM Model:} We consider a model where the interaction between vacuum energy and DM is mediated by the coupling term as mentioned in \eqref{eq2}, and we assume three active neutrinos as in the case of standard model. The base parameters set for this model is 
\begin{eqnarray*}
\label{P1}
P = \{\omega_{\rm b}, \, \omega_{\rm dm},  \, \theta_{s}, \, A_{s}, \,  
n_s, \, \tau_{\rm reio},  \, \xi \}.
\end{eqnarray*}
Here the first six parameters are the baseline parameters of the standard $\Lambda$CDM model \cite{Planck2018}.\\

\noindent\textbf{$\nu$I$\Lambda$CDM Model:} We consider the I$\Lambda$CDM Model with the effective neutrinos and mass number of relativistic species as a free parameter, i.e, 
as free parameters, that is, I$\Lambda$CDM$+N_{\rm eff}+\sum m_{\nu}$. The base parameters set for this mode is therefore
\begin{eqnarray*}
\label{P1}
P = \{\omega_{\rm b}, \, \omega_{\rm dm}, \, \theta_{s}, \, A_{s}, \,  
n_s, \, \tau_{\rm reio},\, \xi, \, N_{\rm eff}, \, \sum m_{\nu}\}.
\end{eqnarray*}
We adopt the evolution of linear perturbations as described in \cite{sk19}. 

\section{Observational constraints and discussions} \label{sec3}
We analyze the I$\Lambda$CDM and $\nu$I$\Lambda$CDM models in contrast with their respective counterparts $\Lambda$CDM and $\nu\Lambda$CDM. We also present the analyses of $w$CDM and $\nu w$CDM models in contrast with the I$\Lambda$CDM and $\nu$I$\Lambda$CDM models. In our analyses, we use the following data sets.
(i) Planck: Planck-2018 \cite{Planck2018} CMB temperature and polarization data comprising of the low-$l$ temperature and polarization likelihoods at $l \leq 29$, temperature (TT) at $l \geq 30$, polarization (EE) power spectra, and cross correlation of temperature and  polarization (TE), also including the Planck-2018 CMB lensing power spectrum likelihood \cite{Planck2018:GL}, (ii) BAO: the latest BAO measurements from SDSS collaboration compiled in Table 3 in \cite{sdss20} (iii) JLA: the compilation of Joint Light-curve Analysis (JLA) supernova Ia data \cite{jla14}  (iv) KiDS: the measurements
of the weak gravitational lensing shear power spectrum
from the Kilo Degree Survey \cite{kids450}, and (iv) R19: the new local value of Hubble constant $H_0 =74.03 \pm 1.42$  km s${}^{-1}$ Mpc${}^{-1}$ \cite{R19} from HST. First, we analyze the I$\Lambda$CDM and $\nu$I$\Lambda$CDM models with Planck+R19 data in order to explore the features of interaction in the dark sector. In the recent previous studies \cite{sk19,122,126}, the $H_0$ value from Planck data alone is reported high enough to be compatible with local $H_0$ measurements. So here we combine Planck and R19 to improve the error bars inline with the recent studies. Likewise, as reported in \cite{sk19}, the $H_0$ value from KiDS data alone is high enough to be compatible with local $H_0$ measurements, and thus there is no tension with the combined Planck+KiDS+R19 data. Also, the constraints on interaction model from Planck+KiDS+HST data are already reported in \cite{sk19}. In the present study, we further add the  BAO+JLA data  to break any possible degeneracy among the parameters and improve the errors bars. Thus, here we assess the I$\Lambda$CDM and $\nu$I$\Lambda$CDM models by considering two data combinations: Planck+R19 and Planck+BAO+JLA+KiDS+R19.  We present these analyses in a comprehensive way while doing a comparative analysis of the six models under consideration: I$\Lambda$CDM, $\nu$I$\Lambda$CDM, $\Lambda$CDM, $\nu\Lambda$CDM, $w$CDM and $\nu w$CDM.

We use the publicly available Boltzmann code \texttt{CLASS} \cite{class} with the parameter inference code \texttt{Monte python+MultiNest} \cite{monte,feroz,buchner} to obtain correlated Monte Carlo Markov Chain (MCMC) samples. We analyze the MCMC smaples using the python package \texttt{GetDist}. In all analyses performed here, we use uniform priors on the model parameters: $\omega_b\in[0.018,0.024]$, $\omega_{\rm cdm}\in[0.10,0.14]$, $100\,\theta_{s}\in[1.03,1.05]$, $\ln(10^{10}A_s)\in[3.0,3.18]$, $n_s\in[0.9,1.1]$, $\tau_{\rm reio}\in[0.04,0.125]$, $\xi\in[-1,1]$, $w_{\rm de}\in[-3,1]$, $N_{\rm eff}\in[1,5]$, and $\sum m_{\nu}\in[0,1]$.

\begin{table*}[hbt!]
\footnotesize
\caption{Constraints (68\% and 95\% CLs)  on  the free and some derived parameters of the I$\Lambda$CDM, $\nu$I$\Lambda$CDM, $\Lambda$CDM, $\nu\Lambda$CDM, $w$CDM and $\nu w$CDM models from Planck+R19 and Planck+BAO+JLA+KiDS+R19 data. The parameter $H_{\rm 0}$ is measured in the units of km s${}^{-1}$ Mpc${}^{-1}$, and $\sum{m_{\nu}}$ in eV (95\% CL). }
\label{tableI}
\setlength\extrarowheight{2pt}
\begin{tabular} { |l| l| l| l| l|l|l|  }  \hline 
 \multicolumn{1}{|c|}{} &\multicolumn{2}{|c|}{Planck+R19}& \multicolumn{2}{|c|}{Planck+BAO+JLA+KiDS+R19} \\
 \hline
 Parameter &  I$\Lambda$CDM     & $\nu$I$\Lambda$CDM  &  I$\Lambda$CDM     & $\nu$I$\Lambda$CDM        \\
 &  \textcolor{blue}{$\Lambda$CDM}     & \textcolor{blue}{$\nu\Lambda$CDM}  &  \textcolor{blue}{$\Lambda$CDM}     & \textcolor{blue}{$\nu\Lambda$CDM}       \\
 &  \textcolor{magenta}{$w$CDM}     & \textcolor{magenta}{$\nu w$CDM}  &  \textcolor{magenta}{$w$CDM}     & \textcolor{magenta}{$\nu w$CDM}       \\
\hline

$10^{2}\omega_{\rm b }$ 

&   $2.234^{+0.013+0.026}_{-0.013-0.025}   $  & $2.220^{+0.024+0.048}_{-0.024-0.044}            $ &   $2.239^{+0.013+0.025}_{-0.013-0.024}   $  & $2.242^{+0.016+0.032}_{-0.016-0.031}            $ \\[1ex]

& \textcolor{blue}{$2.253^{+0.013+0.025}_{-0.013-0.026}$}    &  \textcolor{blue}{$2.276^{+ 0.017+0.033}_{-0.017-0.035}$}
& \textcolor{blue}{$2.268^{+0.012+0.024}_{-0.012-0.023}$}    &  \textcolor{blue}{$2.277^{+ 0.012+0.024}_{-0.012-0.024}$}
\\[1ex]

& \textcolor{magenta}{$2.242^{+0.013+0.025}_{-0.013-0.025}$}    &  \textcolor{magenta}{$2.222^{+ 0.021+0.042}_{-0.021-0.042}$}
& \textcolor{magenta}{$2.244^{+0.012+0.023}_{-0.012-0.024}$}    &  \textcolor{magenta}{$2.247^{+ 0.017+0.033}_{-0.017-0.033}$}
\\[1ex]

\hline
$\omega_{\rm cdm }$  
& $0.1208^{+0.0012+0.0024}_{-0.0012-0.0024}$  &  $0.1191^{+0.0024+0.0049}_{-0.0024-0.0047}$  
& $0.1204^{+0.0011+0.0021}_{-0.0011-0.0021}$  &  $0.1209^{+0.0018+0.0033}_{-0.0018-0.0036}$   
\\[1ex]
 
& \textcolor{blue}{$0.1184^{+0.0011+0.0021}_{-0.0011-0.0021}$}   &  \textcolor{blue}{$0.1223^{+0.0023+0.0046}_{-0.0023-0.0045}$}  
& \textcolor{blue}{$0.1167^{+0.0008+0.0025}_{-0.0008-0.0024}$}   &  \textcolor{blue}{$0.1218^{+0.0018+0.0030}_{-0.0015-0.0032}$}  
\\[1ex]

& \textcolor{magenta}{$0.1197^{+0.0011+0.0022}_{-0.0011-0.0023}$}   &  \textcolor{magenta}{$0.1177^{+0.0024+0.0047}_{-0.0024-0.0046}$}  
& \textcolor{magenta}{$0.1196^{+0.0010+0.0020}_{-0.0010-0.0019}$}   &  \textcolor{magenta}{$0.1203^{+0.0019+0.0038}_{-0.0019-0.0038}$}  
\\[1ex]

\hline

$100 \theta_{s }$ 
& $1.04184^{+0.00029+0.00058}_{-0.00029-0.00058}$ & $1.04211^{+0.00045+0.00089}_{-0.00045-0.00089}$ 
& $1.04186^{+0.00028+0.00055}_{-0.00028-0.00053}$ & $1.04179^{+0.00037+0.00073}_{-0.00037-0.00072}$ 
\\[1ex]

&  \textcolor{blue}{ $1.04208^{ +0.00029+0.00059}_{-0.00029-0.00056}$ }   & \textcolor{blue}{$1.04148^{ +0.00041+0.00080}_{-0.00041-0.00080} $}    
&  \textcolor{blue}{ $1.04226^{ +0.00028+0.00057}_{-0.00028-0.00056}$ }   & \textcolor{blue}{$1.04150^{ +0.00035+0.00069}_{-0.00035-0.00066} $}    
\\[1ex]

&  \textcolor{magenta}{ $1.04192^{ +0.00029+0.00055}_{-0.00029-0.00059}$ }   & \textcolor{magenta}{$1.04227^{ +0.00046+0.00095}_{-0.00046-0.00088} $}    
&  \textcolor{magenta}{ $1.04191^{ +0.00028+0.00055}_{-0.00028-0.00056}$ }   & \textcolor{magenta}{$1.04182^{ +0.00040+0.00081}_{-0.00040-0.00078} $}    
\\[1ex]

\hline
$\ln10^{10}A_{s }$ 
&    $3.058^{+0.013+0.030}_{-0.015-0.028}   $ & $3.051^{+0.016+0.034}_{-0.017-0.031}   $ 
&    $3.055^{+0.013+0.027}_{-0.015-0.026}   $ & $3.056^{+0.014+0.030}_{-0.016-0.029}   $ 
\\[1ex]

&  \textcolor{blue}{$3.054^{+0.014+0.028}_{-0.014-0.027}$}   & \textcolor{blue}{$3.065^{+ 0.016+0.034}_{-0.016-0.030}$}   
&  \textcolor{blue}{$3.057^{+0.013+0.030}_{-0.016-0.028}$}   & \textcolor{blue}{$3.061^{+ 0.014+0.030}_{-0.015-0.028}$}
\\[1ex]

&  \textcolor{magenta}{$3.045^{+0.012+0.025}_{-0.014-0.024}$}   & \textcolor{magenta}{$3.037^{+ 0.014+0.030}_{-0.017-0.028}$}   
&  \textcolor{magenta}{$3.044^{+0.012+0.024}_{-0.013-0.023}$}   & \textcolor{magenta}{$3.048^{+ 0.013+0.030}_{-0.016-0.028}$}
\\[1ex]
\hline

$n_{s } $ 
&  $0.9630^{+0.0035+0.0069}_{-0.0035-0.0072}   $  &  $0.9568^{+0.0092+0.0190}_{-0.0092-0.0170}  $  
&  $0.9642^{+0.0033+0.0063}_{-0.0033-0.0064}   $  &  $0.9659^{+0.0058+0.0120}_{-0.0058-0.0110}  $  
\\[1ex]

&  \textcolor{blue}{$0.9689^{+0.0032+0.0062}_{-0.0032-0.0064}$}  & \textcolor{blue}{ $0.9796^{+0.0063+0.0130}_{-0.0063-0.0130}  $}    
&  \textcolor{blue}{$0.9723^{+0.0029+0.0056}_{-0.0029-0.0058}$}  & \textcolor{blue}{ $0.9792^{+0.0039+0.0079}_{-0.0039-0.0076}  $}
\\[1ex]

&  \textcolor{magenta}{$0.9662^{+0.0032+0.0064}_{-0.0032-0.0061}$}  & \textcolor{magenta}{ $0.9576^{+0.0080+0.0150}_{-0.0080-0.0160}  $}    
&  \textcolor{magenta}{$0.9660^{+0.0030+0.0057}_{-0.0030-0.0059}$}  & \textcolor{magenta}{ $0.9778^{+0.0058+0.0110}_{-0.0058-0.0110}  $}
\\[1ex]
\hline

$\tau_{\rm reio } $  
&  $0.0599^{+0.0071+0.0160}_{-0.0083-0.0150}   $ & $0.0586^{+0.0072+0.0160}_{-0.0081-0.0150}$ 
&  $0.0586^{+0.0067+0.0140}_{-0.0077-0.0140}   $ & $0.0583^{+0.0068+0.0150}_{-0.0078-0.0140}$ 
\\[1ex]

& \textcolor{blue}{$0.0604^{+0.0070+0.0150}_{-0.0069-0.0140}$}   &  \textcolor{blue}{$0.0611^{+ 0.0071+0.0160}_{-0.0081-0.0150}$} 
& \textcolor{blue}{$0.0636^{+0.0069+0.0160}_{-0.0083-0.0140}$}   &  \textcolor{blue}{$0.595^{+ 0.0068+0.0150}_{-0.0079-0.0140}$} 
\\[1ex]

& \textcolor{magenta}{$0.0545^{+0.0062+0.0140}_{-0.0073-0.0130}$}   &  \textcolor{magenta}{$0.0537^{+ 0.0059+0.0130}_{-0.0075-0.0120}$} 
& \textcolor{magenta}{$0.0540^{+0.0065+0.0130}_{-0.0065-0.0120}$}   &  \textcolor{magenta}{$0.0552^{+ 0.0063+0.0150}_{-0.0080-0.0140}$} 
\\[1ex]
\hline

$\xi$  
& $-0.41^{+0.12+0.22}_{-0.12-0.24}   $     &  $-0.50^{+0.20+0.33}_{-0.17-0.38}   $
& $-0.27^{+0.06+0.12}_{-0.06-0.13}   $     &  $-0.24^{+0.08+0.16}_{-0.08-0.17}   $
\\[1ex]

& \textcolor{blue}{$0$}  &  \textcolor{blue}{$0$}
& \textcolor{blue}{$0$}  &  \textcolor{blue}{$0$}
\\ [1ex]
          
& \textcolor{magenta}{$0$}  &  \textcolor{magenta}{$0$}
& \textcolor{magenta}{$0$}  &  \textcolor{magenta}{$0$}
\\ [1ex]
\hline

$w_{\rm de}^{\rm eff}$  
& $-1.14^{+0.04+0.07}_{-0.04-0.08}$  &  $-1.17^{+0.07+0.11}_{-0.06-0.13}$
& $-1.09^{+0.02+0.04}_{-0.02-0.04}$  &  $-1.08^{+0.03+0.06}_{-0.03-0.06}$
\\ [1ex]
          
& \textcolor{blue}{$-1$}  &  \textcolor{blue}{$-1$}
& \textcolor{blue}{$-1$}  &  \textcolor{blue}{$-1$}
\\ [1ex]

& \textcolor{magenta}{$-1.22^{+0.05+0.09}_{-0.05-0.09}$}  
&  \textcolor{magenta}{$-1.30^{+0.10+0.17}_{-0.07-0.19}$}
& \textcolor{magenta}{$-1.11^{+0.03+0.05}_{-0.03-0.05}$}  
&  \textcolor{magenta}{$-1.12^{+0.05+0.09}_{-0.05-0.10}$}
\\ [1ex]
\hline

$N_{\rm eff}$  
& $3.046         $      &  $2.90^{+0.17+0.35}_{-0.17-0.33}      $
& $3.046         $      &  $3.07^{+0.11+0.20}_{-0.11-0.21}      $
\\[1ex]
          
& \textcolor{blue}{$3.046$}  &  \textcolor{blue}{$3.30^{+0.13+0.27}_{-0.13-0.26}$}
& \textcolor{blue}{$3.046$}  &  \textcolor{blue}{$3.28^{+0.08+0.15}_{-0.08-0.16}$}
\\ [1ex]

& \textcolor{magenta}{$3.046$}  &  
\textcolor{magenta}{$2.87^{+0.15+0.31}_{-0.15-0.30}$}
& \textcolor{magenta}{$3.046$}  
&  \textcolor{magenta}{$3.08^{+0.12+0.21}_{-0.12-0.23}$}
\\ [1ex]
\hline

$\sum{m_{\nu}}$
& $0.06         $      &  $<0.15$
 & $0.06         $      &  $<0.13$           
\\[1ex]

& \textcolor{blue}{$0.06$}  &  \textcolor{blue}{$<0.08$}
& \textcolor{blue}{$0.06$}  &  \textcolor{blue}{$<0.11$}
\\ [1ex]

& \textcolor{magenta}{$0.06$}  &  \textcolor{magenta}{$<0.35$}
& \textcolor{magenta}{$0.06$}  &  \textcolor{magenta}{$<0.25$}
\\ [1ex]
\hline
\hline

$\Omega_{\rm{m} }$   
&  $0.272^{+0.010+0.020}_{-0.010-0.020}   $  &  $0.267^{+0.011+0.023}_{-0.011-0.022}         $ 
&  $0.283^{+0.005+0.010}_{-0.005-0.010}   $  &  $0.284^{+0.006+0.011}_{-0.006-0.011}         $ 
\\[1ex]

& \textcolor{blue}{$0.305^{+ 0.007+0.013}_{-0.007-0.013}$}  & \textcolor{blue}{$0.295^{+0.007+0.015}_{-0.008-0.014}$}    
& \textcolor{blue}{$0.294^{+ 0.004+0.008}_{-0.004-0.008}$}  & \textcolor{blue}{$0.295^{+0.004+0.008}_{-0.004-0.008}$}    
\\[1ex]

& \textcolor{magenta}{$0.259^{+ 0.010+0.021}_{-0.010-0.020}$}  & \textcolor{magenta}{$0.257^{+0.010+0.022}_{-0.011-0.020}$}    
& \textcolor{magenta}{$0.283^{+ 0.005+0.009}_{-0.005-0.009}$}  & \textcolor{magenta}{$0.285^{+0.006+0.012}_{-0.006-0.012}$}    
\\[1ex]
\hline
 
$H_{\rm 0}$ 
&   $72.8^{+1.4+2.8}_{-1.4-2.7}        $ &   $72.9^{+1.4+2.8}_{-1.4-2.7}        $ 
&   $71.2^{+0.6+1.1}_{-0.6-1.1}        $ &   $71.1^{+0.6+1.2}_{-0.6-1.1}        $ 
\\[1ex]
 
& \textcolor{blue}{$68.2^{+0.5+1.0}_{-0.5-1.0} $}  & \textcolor{blue}{$70.3^{+1.0+2.0}_{-1.0-2.0}$} 
& \textcolor{blue}{$69.0^{+0.3+0.6}_{-0.3-0.6} $}  
& \textcolor{blue}{$70.1^{+0.5+0.9}_{-0.5-0.9}$}  
\\[1ex]

& \textcolor{magenta}{$74.3^{+1.4+2.8}_{-1.4-2.8} $}  & \textcolor{magenta}{$74.2^{+1.4+2.6}_{-1.4-2.7}$} 
& \textcolor{magenta}{$71.0^{+0.6+1.1}_{-0.6-1.1} $}  & \textcolor{magenta}{$71.1^{+0.6+1.2}_{-0.6-1.2}$}  
\\[1ex]
\hline

$S_{8}$ 
& $0.657^{+0.048+0.089}_{-0.048-0.092}   $ 
&  $0.624^{+0.068+0.130}_{-0.068-0.130}   $ 
& $0.712^{+0.021+0.041}_{-0.021-0.041}   $ 
&  $0.726^{+0.033+0.061}_{-0.033-0.065}   $
\\[1ex]
  
& \textcolor{blue}{$0.816^{+ 0.012+0.024}_{-0.012-0.024}$}  
& \textcolor{blue}{$0.821^{+0.012+0.025}_{-0.012-0.024}$}  
& \textcolor{blue}{$0.797^{+ 0.009+0.018}_{-0.009-0.018}$}  
& \textcolor{blue}{$0.818^{+0.010+0.021}_{-0.010-0.020}$}
\\[1ex]

& \textcolor{magenta}{$0.809^{+ 0.012+0.023}_{-0.012-0.023}$}  
& \textcolor{magenta}{$0.800^{+0.016+0.030}_{-0.014-0.030}$}  
& \textcolor{magenta}{$0.817^{+ 0.010+0.020}_{-0.010-0.019}$}  
& \textcolor{magenta}{$0.813^{+0.014+0.025}_{-0.012-0.027}$}
\\[1ex]
\hline

\end{tabular}
\end{table*}

\begin{figure*}[hbt!]
\includegraphics[width=8cm]{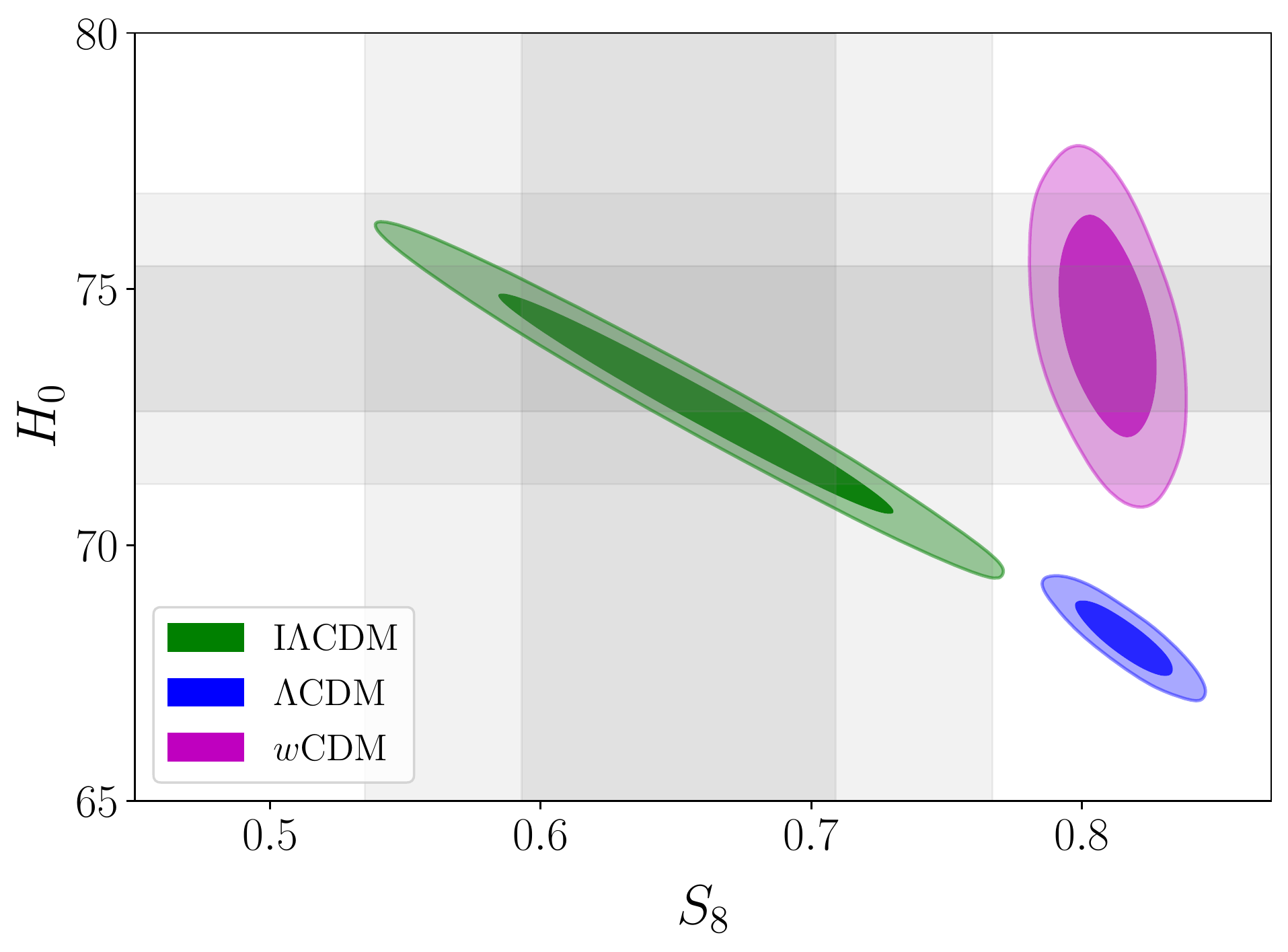}
\includegraphics[width=8cm]{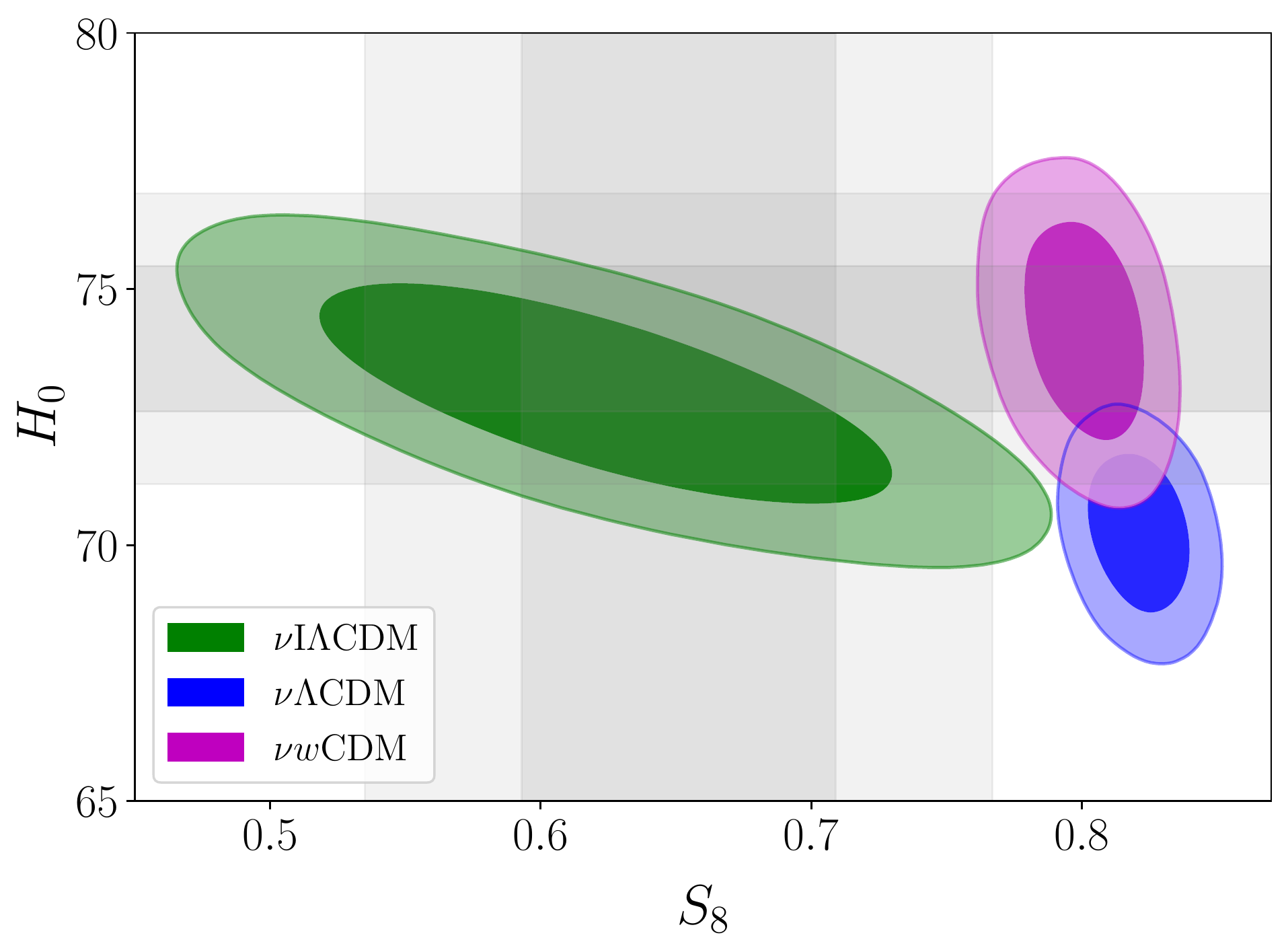}
\includegraphics[width=8cm]{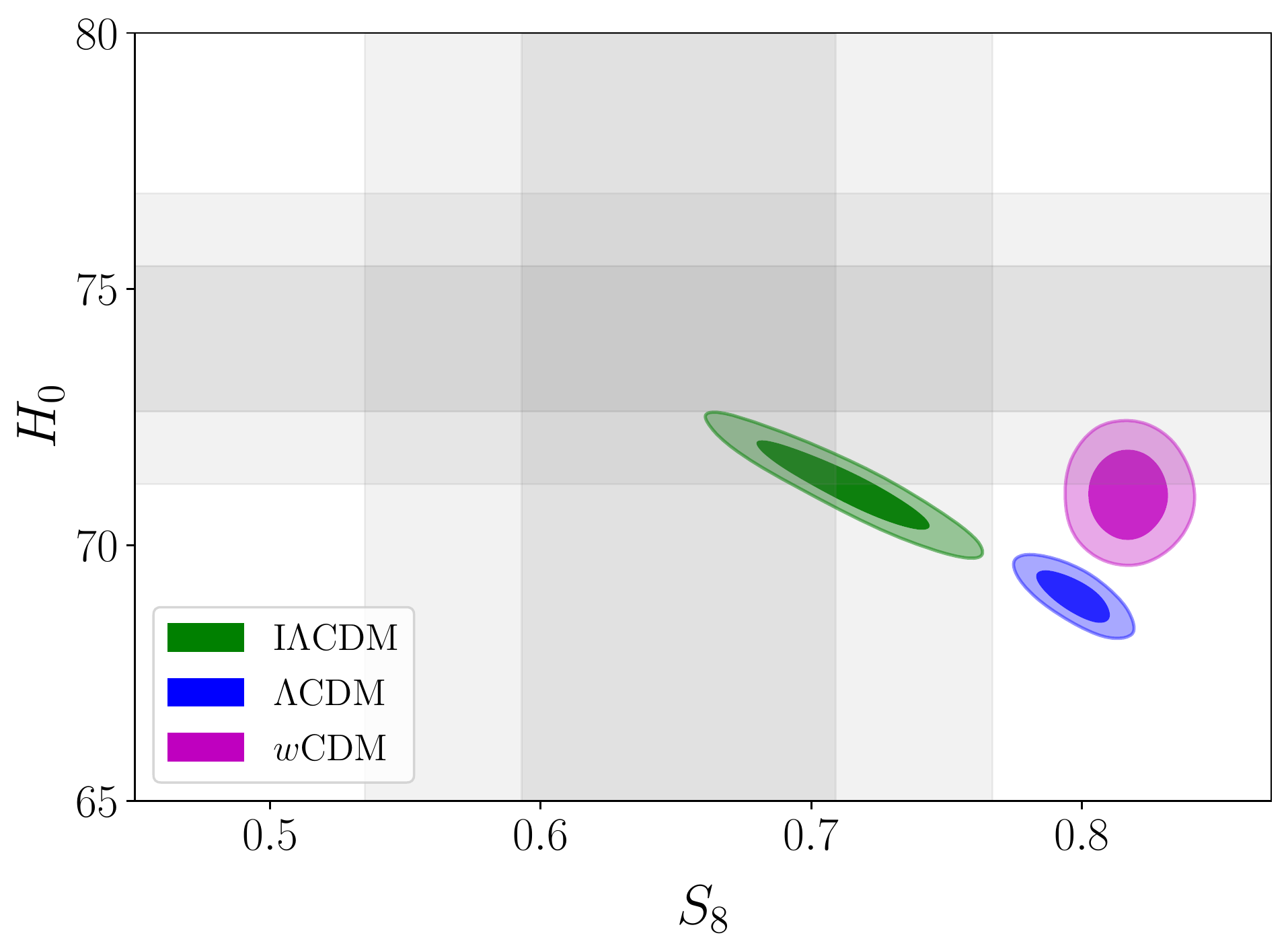}
\includegraphics[width=8cm]{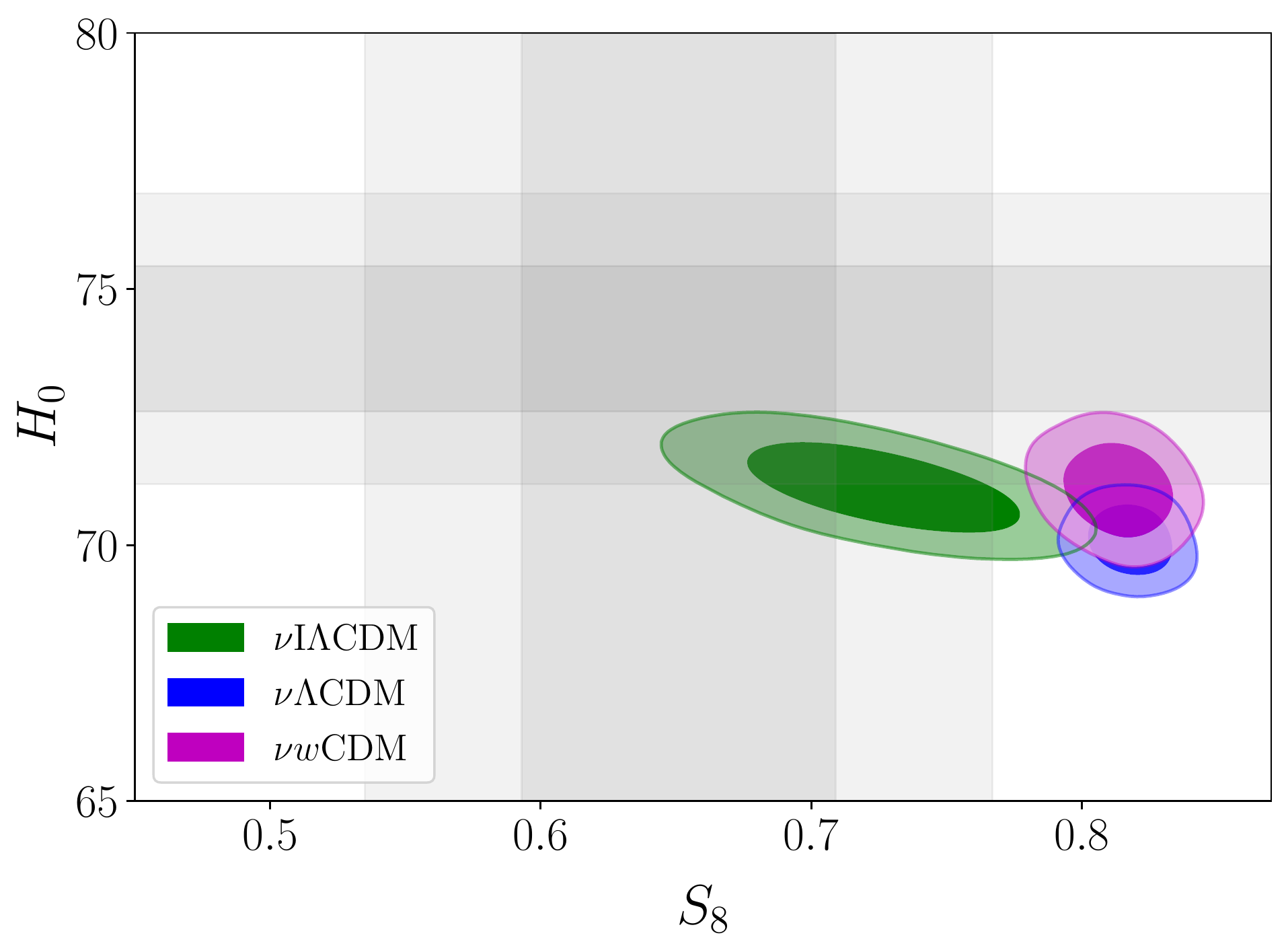}

\caption{2D posteriors for $S_8$ and $H_0$ from Planck+R19 (upper panels) and Planck+BAO+JLA+KiDS+R19 data (lower panels) illustrating how the interacting dark sector can remedy cosmological tensions compared to the other models under consideration. We overlay 2$\sigma$ bands for the
measurements $H_0 =74.03 \pm 1.42$ km s$^{-1}$Mpc$^{-1}$ \cite{R19} and $S_8=0.651\pm 0.058$ \cite{kids450}.}\label{fig2}
\end{figure*}

\begin{figure*}[hbt!]
\includegraphics[width=8.5cm]{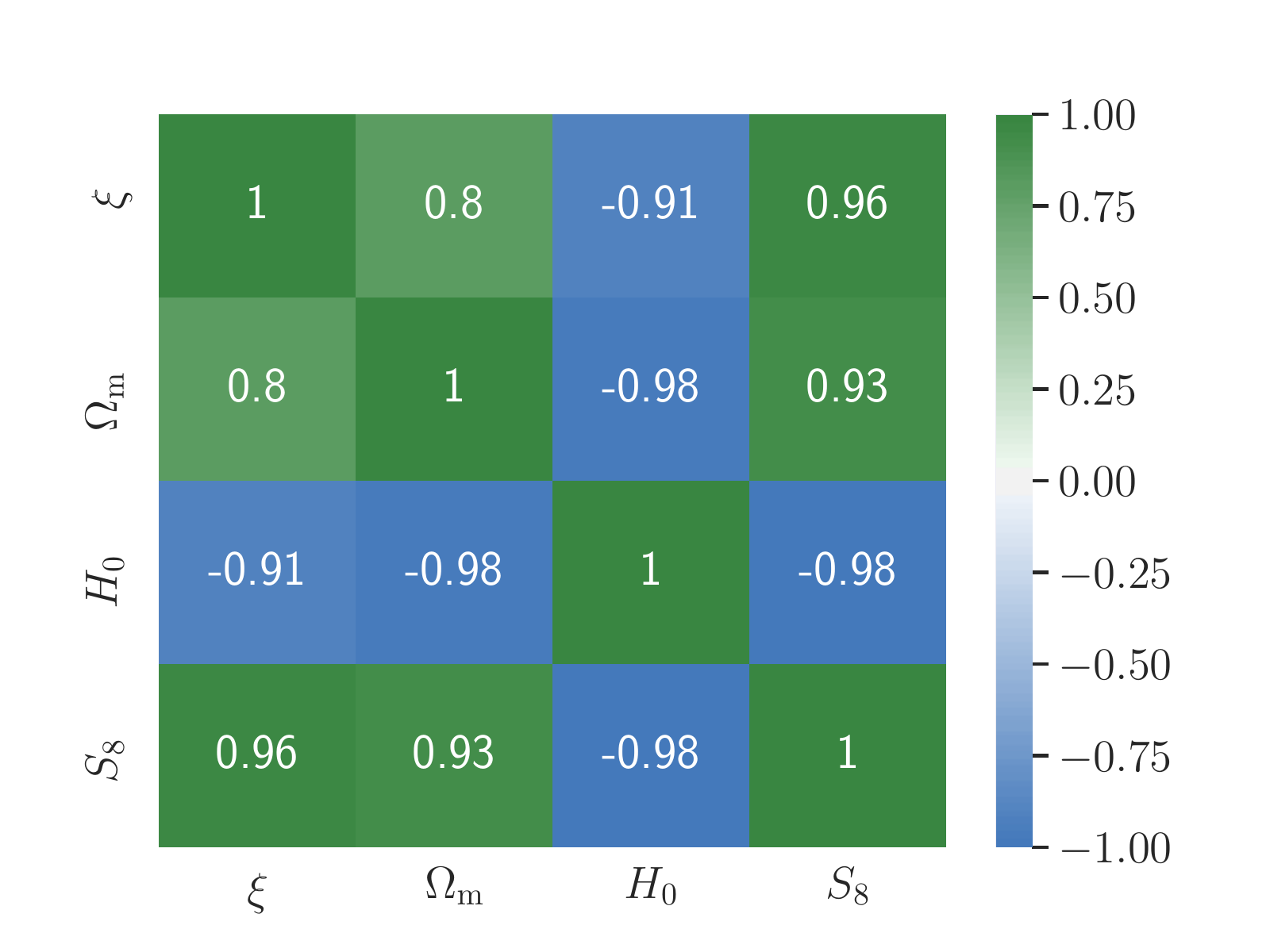}
\includegraphics[width=8.5cm]{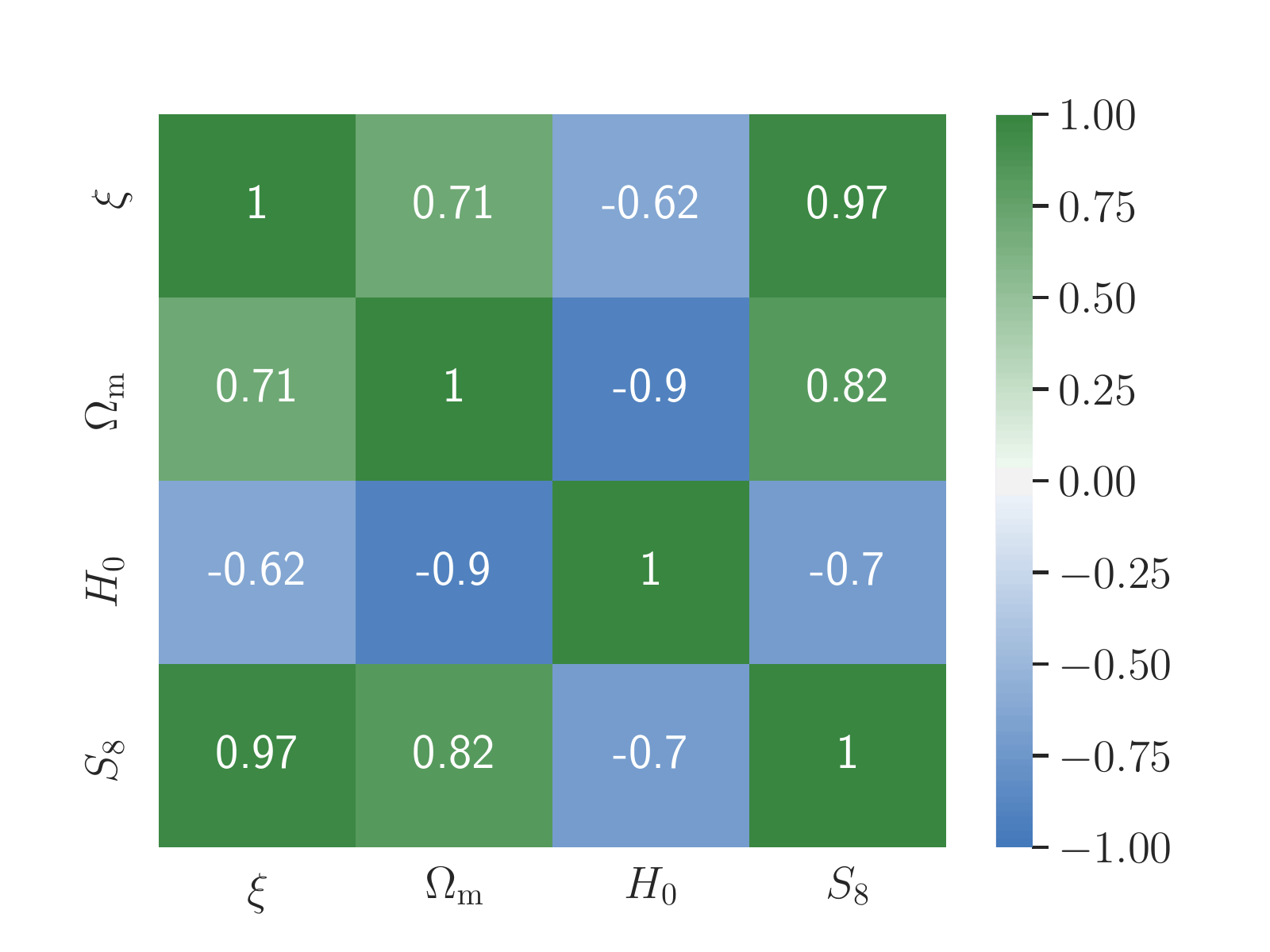}
\includegraphics[width=8.5cm]{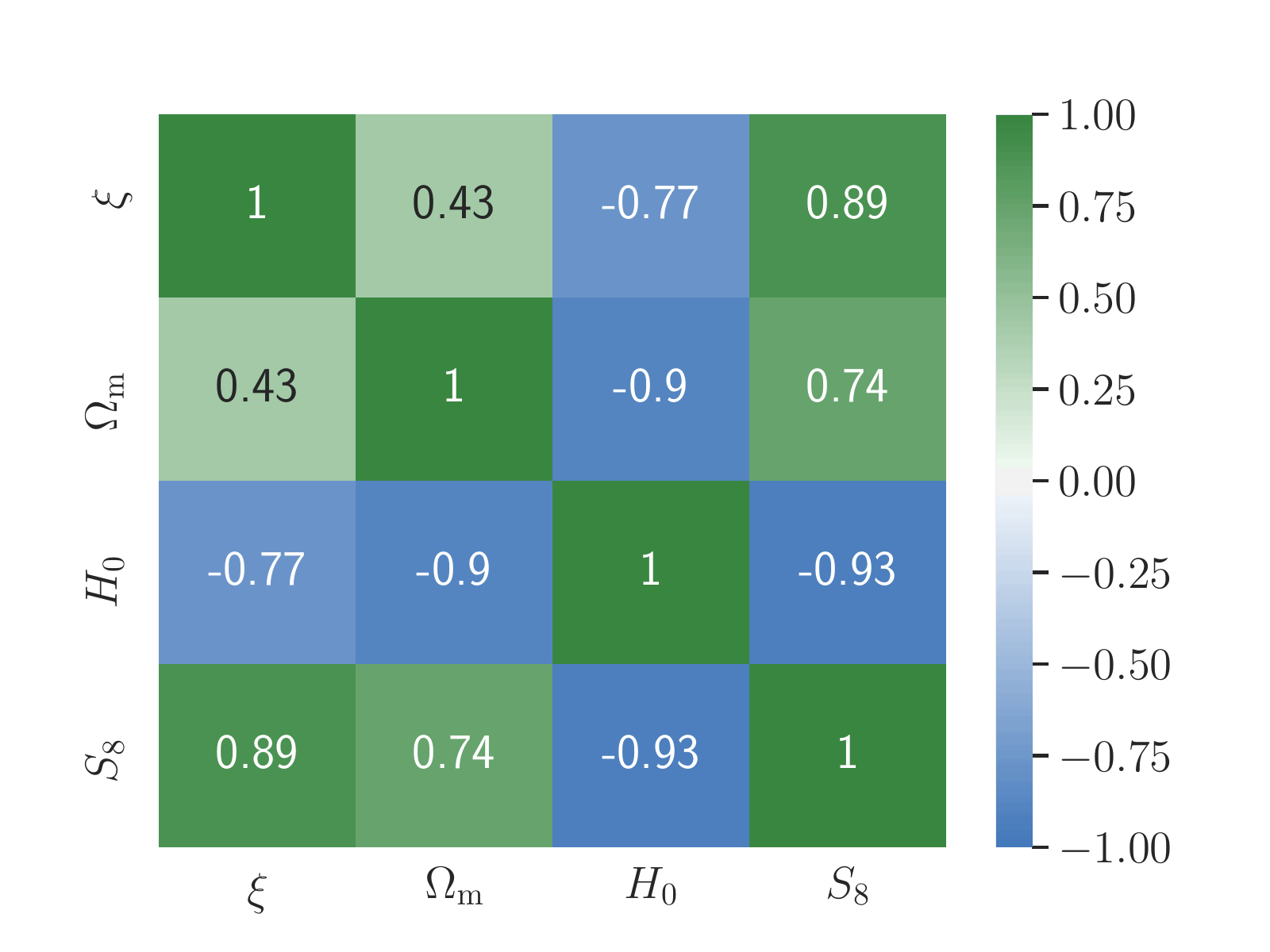}
\includegraphics[width=8.5cm]{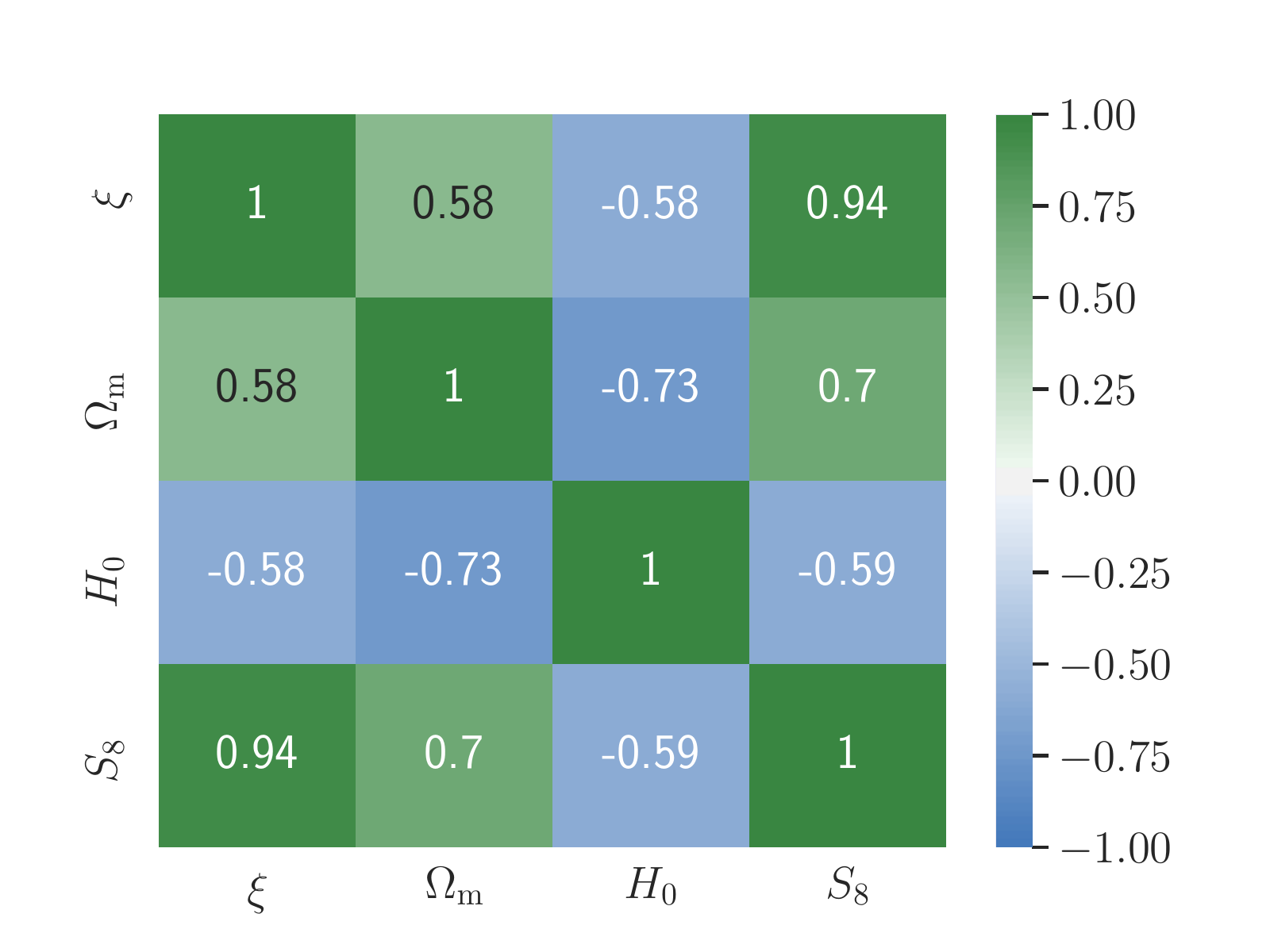}
\caption{Upper-left and upper-right panels respectively show the correlation matrices of some selected parameters of I$\Lambda$CDM and $\nu$I$\Lambda$CDM models  from Planck+R19 data while the lower panles show the corresponding correlation matrices from Planck+BAO+JLA+KiDS+R19 data. }\label{fig3}
\end{figure*}

Table \ref{tableI} summarizes the 68\% and 95\% CL constraints on the free and some derived parameters of the I$\Lambda$CDM and $\nu$I$\Lambda$CDM models in contrast with the parameters of the  $\Lambda$CDM, $\nu\Lambda$CDM, $w$CDM and $\nu w$CDM models, using the Planck+R19 and Planck+BAO+JLA+KiDS+R19 data. We see tight constraints on the parameters of all the six models from the combined data set: Planck+BAO+JLA+KiDS+R19. Also, we observe tight constraints on the parameters of I$\Lambda$CDM, $\Lambda$CDM and $w$CDM models compared to the   $\nu$I$\Lambda$CDM, $\nu\Lambda$CDM and $\nu w$CDM models, as expected, due to the presence of two additional free parameters $\sum{m_{\nu}}$ and $N_{\rm eff}$ in  the later three models. In the following, we analyze  our results to highlight and discuss some key features of the dark sector interaction.\\

\noindent\textbf{(i) Effective phantom-like behavior of DE:} In the I$\Lambda$CDM and $\nu$I$\Lambda$CDM models, we find $\xi = -0.41^{+0.30}_{-0.33}\left(-0.27^{+0.16}_{-0.17}\right)$ and $\xi = -0.50^{+0.44}_{-0.47} \left(-0.24^{+0.20}_{-0.21}\right)$, respectively, both at 99\% CL from the Planck+R19(Planck+BAO+JLA+KiDS+R19) data. Thus, in both these models, the coupling parameter is non-zero and negative, implying the interaction between DM and vacuum energy at 99\% CL wherein the energy transfer takes place from DM to vacuum energy. Further, we find $w_{\rm de}^{\rm eff}=-1.13^{+0.10}_{-0.11}\left(-1.09^{+0.05}_{-0.06}\right)$ and $w_{\rm de}^{\rm eff}=-1.17^{+0.15}_{-0.16}\left(-1.08^{+0.07}_{-0.07}\right)$, both at 99\% CL from the Planck+R19(Planck+BAO+JLA+KiDS+R19) data, for the I$\Lambda$CDM and $\nu$I$\Lambda$CDM models, respectively. Thus, an effective phantom-like behavior of DE prevails in the two models of interaction without ``phantom crossing" at 99\% CL. It should be noted that this effective phantom-like behavior of DE is accompanied by effective positive pressure of DM (see eq.\eqref{eq6}) at 99\% CL for the I$\Lambda$CDM and $\nu$I$\Lambda$CDM models. Also, in our analyses, we obtain phantom behavior of DE in the $w$CDM and $\nu w$CDM models at 99\% CL from the Planck+R19 and Planck+BAO+JLA+KiDS+R19 data.\\

\noindent\textbf{(ii) Relaxing the $H_0$ and $S_8$ tensions:} In Fig. \ref{fig2}, 2D posteriors for $S_8$ and $H_0$ are shown for all the models under consideration where we overlay 2$\sigma$ bands for the
measurements $H_0 =74.03 \pm 1.42$ km s$^{-1}$Mpc$^{-1}$ \cite{R19} and $S_8=0.651\pm 0.058$ \cite{kids450}. One may see that the $H_0$ and $S_8$ tensions are relaxed considerably in the I$\Lambda$CDM and $\nu$I$\Lambda$CDM models constrained with Planck+R19 and Planck+BAO+JLA+KiDS+R19 data. Here, it is interesting to note that the coupling parameter $\xi$ finds strong correlations with $H_0$ and $S_8$, which may be observed from Fig. \ref{fig2}, and more explicitly from the plot of correlation matrices in Fig. \ref{fig3}. The strong negative correlation of $\xi$ with $H_0$ implies that lower values of $\xi$ correspond to higher values of $H_0$. It follows from a natural physical interpretation of the interaction between DM and vacuum energy. For, $\xi<0$ implies that energy of DM is transferred to vacuum energy, and thereby causes a phantom like effect of DE. This in turn causes faster expansion of the late Universe, and hence the higher values of $H_0$ compared to the $\Lambda$CDM model. On the other hand, the $H_0$ tension in the $\Lambda$CDM model is relaxed neither with Planck+R19 data nor with Planck+BAO+JLA+KiDS+R19 data but $H_0$ gets bit larger values in $\nu$I$\Lambda$CDM model due to the larger values of $N_{\rm eff}$. The $H_0$ tension is relaxed in both $w$CDM and $\nu w$CDM models with Planck+R19 as well as Planck+BAO+JLA+KiDS+R19 data because of the phantom behavior of DE in these models. We notice that the $S_8$ tension is not relaxed in any of the four companion models: $\Lambda$CDM, $\nu\Lambda$CDM, $w$CDM and $\nu w$CDM under consideration.\\ 

\begin{figure}[h]
\includegraphics[width=9cm]{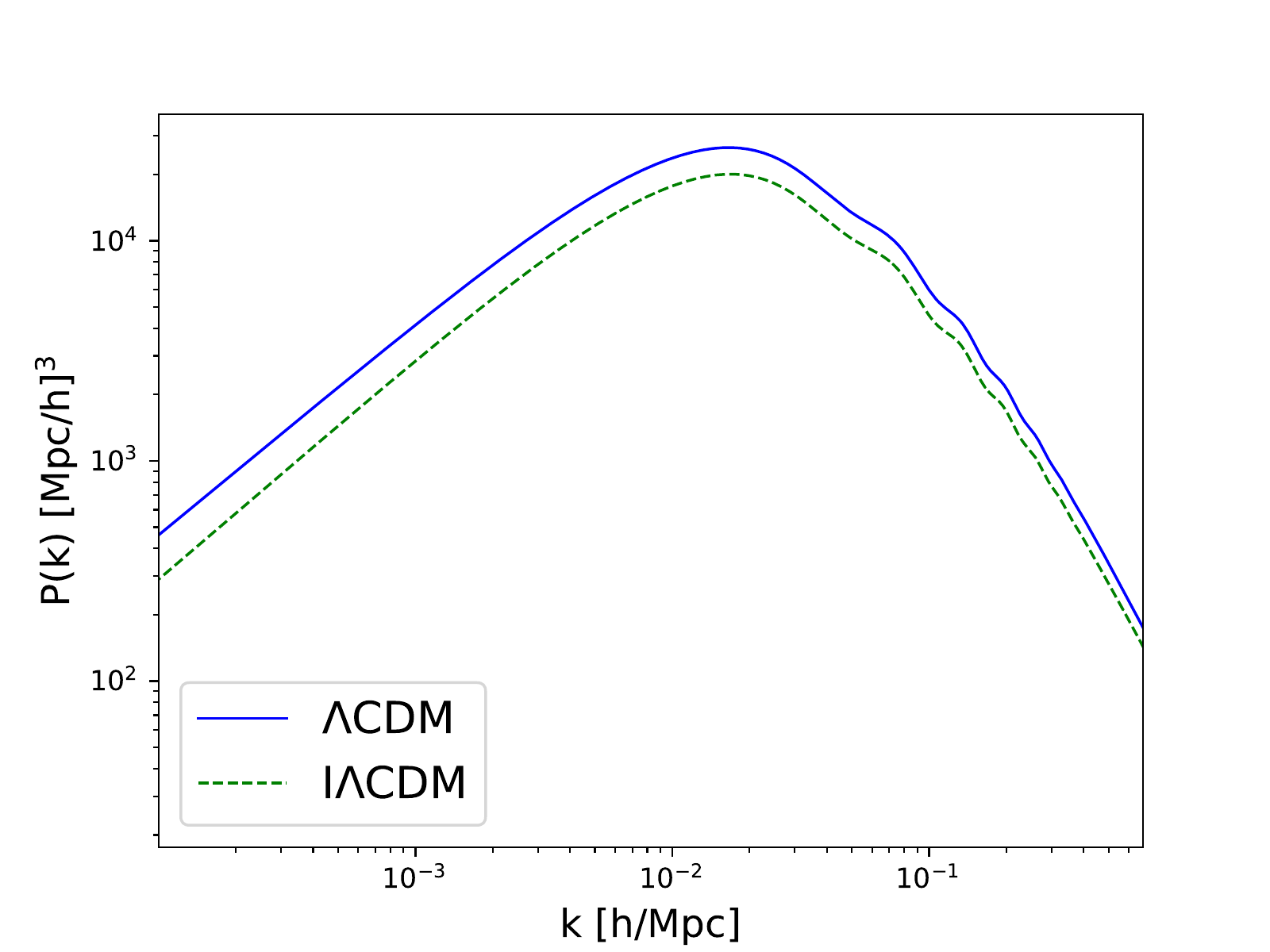}
\caption{Matter power spectra of the I$\Lambda$CDM and $\Lambda$CDM models with the best-fit mean values of the parameters displayed in Table \ref{tableI}.}\label{fig4}
\end{figure}

\noindent\textbf{(iii) A possible solution to the small scale problems:} From Table \ref{tableI}, we notice that the matter density parameter $\Omega_{\rm m}$ and $S_8$ get lower values in I$\Lambda$CDM and $\nu$I$\Lambda$CDM models as compared to $\Lambda$CDM, $\nu\Lambda$CDM, $w$CDM and $\nu w$CDM models. In Fig. \ref{fig4}, we notice less power on all the linear scales in the matter power spectrum of I$\Lambda$CDM model when compared to the $\Lambda$CDM. A plausible physical explanation could be that effective phantom like behavior of DE in the interaction models leads to faster expansion of the late Universe, which dilutes the  matter and decrease the structure formation on linear and non-linear scales. These observations indicate that allowing the interaction among the dark sector components could pave the way for resolving the small scale problems associated with the $\Lambda$CDM model. However, in order to assess the effects of DM-DE coupling on non-linear scales, one needs to perform N-body simulations of the interaction scenario. For instance, the results of a series of high-resolution hydrodynamical N-body simulations of DM-DE coupling scenario are presented in \cite{17,89}, wherein the authors carried out a basic analysis of the formation of non-linear structures, and
found that a larger coupling strength leads to a lower average halo concentration. Further, they deduced that the halo density profiles get shallower in the inner part of massive halos
with increasing value of the coupling, and the halo concentrations at $z=0$ are significantly reduced with respect to
$\Lambda$CDM, proportionally to the value of the coupling. They concluded that such effects alleviate the tensions between observations and the $\Lambda$CDM model on small scales, implying that the coupled DE models could be viable alternatives to the standard $\Lambda$CDM model.

\noindent\textbf{(iv) Consistency with the standard neutrinos and mass number:} From Table \ref{tableI}, we notice that the constraints on $\sum m_{\nu}$ and $N_{\rm eff}$ in the interaction models are consistent with the standard model predictions. This observation is interesting in the sense that higher values $H_0$ usually correspond to higher values of $N_{\rm eff}$ because of the positive correlation between these two parameters \cite{107,109}, leading to larger values of $H_0$. But in the $\nu$I$\Lambda$CDM  model, we find no correlation between $H_0$ and $N_{\rm eff}$. Thus, the $\nu$I$\Lambda$CDM  model yields higher values of $H_0$ and smaller values $S_8$ while being consistent with the standard model predictions of the neutrinos and mass number. We notice that $\nu w$CDM model is also consistent with the standard model predictions of the neutrinos and mass number but this model does not relax the $S_8$ tension.\\

\noindent\textbf{(v) Better fit to the data:} The Bayes' factor $B_{\mathcal{M}_a,\mathcal{M}_b}$ of two competing models $\mathcal{M}_a$ and $\mathcal{M}_b$ is given by \cite{jef95,akarsu19}
\begin{equation}\label{bayes_factor}
B_{\mathcal{M}_a,\mathcal{M}_b} = \frac{\mathcal{E}_{\mathcal{M}_a}}{\mathcal{E}_{\mathcal{M}_b}},
\end{equation}
where $\mathcal{E}_{\mathcal{M}_a}$ and $\mathcal{E}_{\mathcal{M}_b}$ are respectively the Bayesian evidences of the models $\mathcal{M}_a$ and $\mathcal{M}_b$. It is interpreted as per the Jeffery's scale in four different cases depending on the value of $|\ln B_{\mathcal{M}_a,\mathcal{M}_b}|$, viz., values lying in the ranges [0,1), [1,3), [3,5) and $[5,\infty)$ imply the strength of the evidence to be weak, positive, strong and the strongest, respectively. From our analyses, we find\\

\noindent $\ln B_{\rm \Lambda CDM,\rm I\Lambda CDM}=-3.58$, $\ln B_{\rm \Lambda CDM,\rm wCDM}=-8.11$,\\ 
$\ln B_{\rm \nu\Lambda CDM,\rm \nu I\Lambda CDM}=-2.95$, $\ln B_{\rm \nu\Lambda CDM,\rm \nu wCDM}=-7.82$,\\
 
\noindent with Planck+R19 data, and\\ 

\noindent $\ln B_{\rm \Lambda CDM,\rm I\Lambda CDM}=-8.49$, $\ln B_{\rm \Lambda CDM,\rm wCDM}=-6.51$,\\ 
$\ln B_{\rm \nu\Lambda CDM,\rm \nu I\Lambda CDM}=-2.65$, $\ln B_{\rm \nu\Lambda CDM,\rm \nu wCDM}=-1.01$,\\ 

\noindent with Planck+BAO+JLA+KiDS+R19 data. We observe that the I$\Lambda$CDM, $\nu$I$\Lambda$CDM, $w$CDM and $\nu w$CDM models find better fit to the Planck+R19 data when compared to their respective counterparts $\Lambda$CDM and $\nu\Lambda$CDM wherein we observe strong evidence in each case as per the Jeffery's scale. One may notice that the $\Lambda$CDM model does not accommodate/accept the R19 prior in the fit, and hence shows a big tension with the $H_0$ values given by R19 prior. The interaction models find better fit to the Planck+BAO+JLA+KiDS+R19 data compared to the other models under consideration.

It deserves mention that the data combinations considered in this study include the $H_0$ prior, while in some recent studies in the literature, e.g.\cite{prior1, prior2}, it is argued that the use of the $H_0$ prior may not be appropriate in obtaining the unbiased constraints on the model parameters. So we check our results by dropping the $H_0$ prior. Indeed the results with Planck+R19 should not be used for the final conclusions due to the strong degeneracy of $\xi$ with $H_0$. These are given here just to revisit the results in the literature with this combination. So our main and new results in this study are with the combined data: Planck+BAO+JLA+KiDS+R19. After dropping the $H_0$ prior from this combination, we do not find significant changes in the interaction model parameters. For instance, at 99\% CL, we find $\xi=-0.22^{+0.17}_{-0.18}$ and $H_0=70.6\pm1.6$ km s$^{-1}$Mpc$^{-1}$ in the I$\Lambda$CDM model, and $\xi=-0.20^{+0.20}_{-0.22}$ and $H_0=70.6\pm1.6$ km s$^{-1}$Mpc$^{-1}$ in the $\nu$I$\Lambda$CDM model.  

\section{Comparison with previous studies}\label{sec4}
This study is of course an update of our previous work \cite{sk19} wherein we reported constraints on I$\Lambda$CDM model compared to the $\Lambda$CDM model using Planck 2015, HST and KiDS data. The constraints on the six models under consideration here are recently reported in \cite{122} with the Planck, Planck+BAO and Planck+R19 data. So our results are directly comparable with the results reported in \cite{122}. Indeed, here we have revisited the results of \cite{122} with Planck+R19 data while we have reported much improved constraints on the parameters with Planck+BAO+JLA+KiDS+R19 data compared to the Planck, BAO and R19 data constraints reported in \cite{122}, in all the models. We have the following new results/developments in the present study.

(i) We have explicitly shown and quantified the phantom-like behavior of effective DE in the interaction models without actual ``phantom crossing". We have found non-zero coupling parameter at 99\% CL with Planck+R19 data, as also reported in \cite{122}. Here, we have further found that this result remains true with the combined Planck+BAO+JLA+KiDS+R19 data with improved constraints.

(ii)  Relaxation of only $H_0$ tension is reported in the interaction models in \cite{122}. On the other hand, we have reported that both $H_0$ and $S_8$ tensions are relaxed with the phantom-like behavior of effective DE in the interaction models wherein we have displayed explicit correlation matrices of the parameters, and explained how the two tensions are relieved simultaneously in the interaction models with the phantom-like behavior of effective DE. It is interesting to note that the $H_0$ and $S_8$ tensions are relaxed significantly in the interaction models with the combined Planck+BAO+JLA+KiDS+R19 data as well and improved constraints.

(iii) We have noticed less power on all the linear scales in the matter power spectrum of I$\Lambda$CDM model when compared to the $\Lambda$CDM. These observations are interesting and indicate that allowing the interaction among the dark sector components could pave the way for resolving the small scale problems associated with the $\Lambda$CDM model. However, this should be considered a hand-waving result from this study because it requires a detailed investigation with specific methods in order to properly assess the effects of DM-DE coupling on non-linear scales, which is beyond the scope of this paper.

(iv) Here, $\nu$I$\Lambda$CDM model shows consistency with the standard effective neutrino mass and number with Planck+R19 data, also reported in \cite{122}. We have further found that this result remains true with the combined Planck+BAO+JLA+KiDS+R19 data with improved constraints.

(v) We have calculated the Bayesian evidence for statistical fit of each model explicitly using the \texttt{MultiNest} code, and found that the interaction models find better fit to the Planck+BAO+JLA+KiDS+R19 data compared to the other models under consideration. In \cite{122}, Bayesian evidence of the fit is not reported. Also, to our knowledge, the Bayesian evidence of the statistical fit of the vacuum energy interaction  models is not reported in other studies using the \texttt{MultiNest} code.

We have compared the results of the interaction models with the other companion models under consideration. Overall, the new features and improved constraints on the interaction models in this study could be interesting and worthwhile for future studies in this direction.

\section{Final Remarks}\label{sec5}
We have used the Planck 2018, BAO, JLA, KiDS and HST data to investigate two extensions of the base $\Lambda$CDM model, viz.,  I$\Lambda$CDM and $\nu$I$\Lambda$CDM models, wherein the non-minimal interaction strength between vacuum energy and CDM is proportional to the vacuum energy density and expansion rate of the Universe. We have compared the results of these models with the companion $\Lambda$CDM, $\nu\Lambda$CDM, $w$CDM and $\nu w$CDM models in order to show some features and consequences of the phenomenological interaction in the dark sector by considering two combinations of data sets, namely, Planck+R19 and Planck+BAO+JLA+KiDS+R19. In both the interaction models, we have found non-zero coupling in the dark sector up to 99\% CL with energy transfer from dark matter to vacuum energy, and observed a phantom-like behavior of the effective DE without actual ``phantom crossing". The well-known tensions on the cosmological parameters $H_0$ and $\sigma_8$, prevailing within the $\Lambda$CDM cosmology, are relaxed significantly in these models wherein the $\nu$I$\Lambda$CDM model shows consistency with the standard effective neutrino mass and number. The $\nu w$CDM model is also found to be consistent with the standard model predictions of the neutrinos and mass number but this model does not relax the $S_8$ tension. We have observed phantom-like behavior of DE and the relaxation of $H_0$ tension in the $w$CDM and $\nu w$CDM models. We have noticed that interaction in the dark sector could possibly pave the way to resolving the small scale problems associated with the $\Lambda$CDM model. Both the interaction models find a better fit to the combined data when compared to the companion $\Lambda$CDM, $\nu\Lambda$CDM, $w$CDM and $\nu w$CDM models. Overall, in the context of the cosmological tensions and fit to the observational data, we have found that the interaction models do a better job than the companion models under consideration. However, the choice of the coupling term in these interaction models is purely phenomenological for which a natural guidance from the fundamental physics does not exist, and therefore it could be worthwhile to investigate the dark sector interaction using some model-independent approach such as the Gaussian process (e.g. \cite{c1,c2} and references therein). Such a model-independent approach could also be useful to measure $H_0$. For instance, in a recent study \cite{c21}, a model-independent estimate of $H_0$ is obtained using the Gaussian process with SN+CC+BAO+H0LiCOW data, viz., $H_0=73.78\pm0.84$ km s${}^{-1}$ Mpc${}^{-1}$, which is consistent with $H_0 =74.03 \pm 1.42$  km s${}^{-1}$ Mpc${}^{-1}$, the measurement from HST \cite{R19}. However, without H0LiCOW data (known to be consistent with R19), the obtained $H_0$ values are in small tension with R19.  Finally, it is worth mentioning that the interaction among dark sector components of the Universe could have very interesting and far reaching consequences in the contemporary particle physics and  cosmology. Since physics of the dark sector components is not yet well-understood, the phenomenological studies of interaction in the dark sector, like the ones carried out in the present work, can reveal interesting and worthwhile features and consequences of the interaction in the dark sector, which in turn  could possibly pave the way to theoretical and experimental progress in this direction. In the next decade or so, the cosmology community is gearing up for solution to various problems associated with the $\Lambda$CDM model such as the Hubble constant tension and others \cite{CI12020,CI22020,CI32020,CI42020}. With the future CMB experiments such as CMB-S4, and the cosmic surveys such as Euclid and LSST, one may expect the $H_0$ estimate within an uncertainty of $\sim 0.15\%$. Indeed, a coordinated effort is required by the community in the coming decade to test the $\Lambda$CDM model and other exotic cosmologies wherein the next-generation experiments are likely to usher in a new era of cosmology.

\begin{center}
\textbf{Acknowledgments}
\end{center}

The author thanks to the referee for valuable comments and suggestions which helped to improve the clarity and quality of some results in this study. The author thanks to Rafael C. Nunes and \"{O}zg\"{u}r Akarsu for fruitful discussions on this work, and acknowledges the support from SERB-DST project No. EMR/2016/000258.

\end{document}